\documentstyle[12pt,epsf,amsfonts,amssymb]{article}

\def\a{\alpha}
\def\b{\beta}
\def\ben{\begin{equation}}
\def\een{\end{equation}}
\def\bena{\begin{eqnarray}}
\def\eena{\end{eqnarray}}

\def\N{{\cal N}}
\def\O{{\cal O}}

\def\mr{{\mathbb R}}
\renewcommand{\c}{{\rm const.}}

\newcommand{\bome}{\mbox{\boldmath $\omega$}}
\newcommand{\I}{{\mathcal I}}
\renewcommand{\H}{{\mathcal H}}

\newcommand{\Q}{{\mathcal Q}}

\newcommand{\bth}{\mbox{\boldmath $\theta$}}
\begin{document}

\begin{titlepage}
\bigskip
\rightline{}
\rightline{gr-qc/....}
\bigskip\bigskip\bigskip\bigskip
\centerline
{\Large \bf {Stability in Designer Gravity}}
\bigskip\bigskip
\bigskip\bigskip

\centerline{\large Thomas 
Hertog\footnote{hertog@vulcan.physics.ucsb.edu} and 
Stefan Hollands$^{1,}$\footnote{hollands@theorie.physik.uni-goettingen.de}}
\bigskip\bigskip
\centerline{$^1$\em Department of Physics, UCSB, Santa Barbara, CA 93106}
\bigskip
\centerline{$^2$\em Inst. f. Theor. Physik, Georg-August-Universit\"at, D-37077 G\"ottingen}
\bigskip\bigskip

\begin{abstract}

We study the stability of designer gravity theories, in which one considers gravity coupled to a tachyonic scalar 
with anti-de Sitter boundary conditions defined by a smooth function $W$. We construct Hamiltonian generators of the 
asymptotic symmetries using the covariant phase space method of Wald et al. and find they differ from the spinor 
charges except when $W=0$. The positivity of the spinor charge is used to establish a lower bound on the conserved 
energy of any solution that satisfies boundary conditions for which $W$ has a global minimum. A large class of 
designer gravity theories therefore have a stable ground state, which the AdS/CFT correspondence indicates should be 
the lowest energy soliton. We make progress towards proving this, by showing that minimum energy solutions are static.
The generalization of our results to designer gravity theories in higher dimensions 
involving several tachyonic scalars is discussed.

\end{abstract}

\end{titlepage}

\baselineskip=18pt

\setcounter{equation}{0}
\section{Introduction}

In theories of gravity coupled to matter, the theory is usually fully determined by the action. The boundary 
conditions at infinity are often not independent, but uniquely determined by basic requirements such as finite 
total energy. This is not the case, however, for gravity coupled to certain scalar field theories in 
asymptotically anti-de Sitter (AdS) spacetimes. In particular, it has been shown that scalars $\phi$ with mass 
$m^2$ at or slightly above the Breitenlohner-Freedman (BF) bound \cite{Breitenlohner82} admit a large class of 
boundary conditions, defined by an essentially arbitrary real function 
$W$, which prescribes the relation between the asymptotic value 
of the scalar field and its normal derivative at infinity. Furthermore, for
each choice of $W$, one can derive manifestly finite 
expressions for the total energy and angular momentum of the 
theory \cite{Hertog05}.

Theories of this type were called designer gravity theories, since their properties depend 
significantly on the choice of boundary conditions. By choosing the appropriate boundary 
condition function $W$ one can, for example, specify the 
number as well as the masses of spherical soliton solutions. 
It was also found that certain boundary conditions admit 
AdS-black holes with scalar hair \cite{Hertog04,Martinez04,Hertog05b}. 
Designer gravity theories therefore provide a
useful context in which one can study what are the requirements 
on scalar matter for the no hair theorems to hold.

Interesting examples of designer gravity theories arise in string theory compactifications on a sphere. String theory with $AdS_4 \times S^7$ boundary 
conditions, for instance, reduces at low energies to ${\cal N}=8$ gauged 
supergravity in four dimensions, which contains several $m^2=-2$ scalars.
It is an obvious question which---if any---choices of $W$ would yield a stable theory, i.e., one 
whose energy is bounded from below. Furthermore, in such a theory, it would 
be of interest to classify all solutions that locally minimize the energy.  

Insight into these questions can be gained from the AdS/CFT correspondence
\cite{Maldacena98}. This states that, with $W=0$ 
boundary conditions, the gravity theory is equivalent to a 2+1 conformal field 
theory (CFT) that is defined on the boundary spacetime. Adopting more 
general $W \neq 0$ boundary conditions on one (or several) 
of the tachyonic scalars corresponds to the addition of a potential 
term $\int W({\cal O})$ to the dual CFT action, where ${\cal O}$ is the 
field theory operator that is dual to the bulk scalar \cite{Witten02, Berkooz}.
Based on the AdS/CFT duality, it was conjectured in \cite{Hertog05} that 
(a) there is a lower bound on the gravitational energy in those designer
gravity theories where $W$ is bounded from below and that
(b) the solutions locally minimizing the energy are given by the spherically 
symmetric, static soliton configurations found in~\cite{Hertog05}.

In this paper, we make progress towards proving these conjectures. 
First, we prove that if the function $W$ is bounded from below, 
then so is the energy of the gravity theory. In fact, we present a lower 
bound on the gravitational energy in terms of the global minumum of $W$, 
see eq.~(\ref{bound}). Second, we prove that, under certain 
technical assumptions, local minima of the energy must be static, 
i.e., admit a globally defined surface 
orthogonal timelike Killing field. However, we have not yet been able to 
fully characterize these minima. In particular, we do not know whether 
they are given by the spherical soliton solutions 
found in~\cite{Hertog05}, or whether
they encompass a wider class of solutions. Progress towards these problems
can perhaps be made most easily by numerical methods.

Our proofs of these statements only involve arguments in the classical 
gravity theory, and do not rely on the AdS/CFT correspondence. However, 
the lower bound ~(\ref{bound}) agrees with the lower bound on the energy of
dual field theory configurations. Our result can therefore be viewed as a 
test of the AdS/CFT correspondence.

We also outline the generalization of our results to other designer 
gravity theories in higher dimensions involving several tachyonic 
scalar fields.

Our {\bf notations and conventions} are the same as in~\cite{waldbook}. In particular, indices on tilded tensor fields
$\tilde t_{abc\dots}$ are raised and lowered with the unphysical metric $\tilde g_{ab}$ and its inverse, whereas 
indices on untilded fields are raised and lowered with the physical metric $g_{ab} = \Omega^{-2} \tilde g_{ab}$, and 
its inverse. We set the AdS radius and $8\pi G$ equal to one.

\setcounter{equation}{0}
\section{Asymptotics and Conserved Charges} 

Most generally, we are interested in theories of gravity coupled to one or more scalar fields with potential 
$V(\phi)$. In particular, we will consider potentials that have a critical point $\phi_0$ where $V(\phi_0) <0$ and 
where certain small scalar fluctuations are tachyonic, $m^2 <0$, with the scalar mass in the range
\ben\label{range}
m^2_{BF} \le m^2 < m^2_{BF} +1.
\een
Here $m^2_{BF} =-(d-1)^2/4$ is the Breitenlohner-Freedman (BF) bound \cite{Breitenlohner82}
in $d$ spacetime dimensions.
For definiteness we will focus on the case of ${\cal N}=8$ gauged supergravity in $d=4$ dimensions
and comment on generalizations at the end. This theory can be consistently truncated to include
just gravity and a single scalar field \cite{Duff99}, given by the Lagrangian density 
(viewed as a $4$-form)
\begin{equation}\label{theory}
{\bf L} = \frac{1}{2} \, d^4 x \sqrt{ - g} \, [R - (\nabla \phi)^2 - 2V(\phi)] ,  
\end{equation}
with
\ben\label{pot}
V(\phi) = -2-\cosh \sqrt{2}\phi. 
\een
The potential (\ref{pot}) has a maximum at $\phi=0$. It is unbounded from below, but small 
fluctuations have $m^2 = -2$, which is above the BF bound and within the range (\ref{range}). 
To specify the theory completely, we must also provide a set of boundary conditions, 
to which we now turn.

\subsection{Asymptotic Conditions}

\begin{enumerate}
\item One can attach a boundary, $\I \cong {\mathbb R} \times S^{2}$ 
to $M$ such that $\tilde M = M \cup \I$ is a manifold with boundary. 
\item
On $\tilde M$, there is a smooth\footnote{By ``smooth'', we always mean $C^\infty$.} function 
$\Omega$, with $\Omega = 0$ and $\tilde n_a = \tilde \nabla_a \Omega \neq 0$ on $\I$, such that
\ben \label{metric1}
\tilde g_{ab} = \Omega^2 g_{ab}, \quad \tilde \phi = \Omega^{-1} \phi
\een
are smooth at $\I$. We also require that
\ben\label{metric2}
d\tilde s^2 = d\Omega^2 -dt^2 + d \theta^2 + \sin^2 \theta \, d\varphi^2 
+ \dots \quad {\rm near} \,\, \I,   
\een 
and 
\ben\label{asscalar}
\tilde \phi = \alpha  + \beta \Omega + \dots \quad {\rm near} \,\, \I. 
\een
Here, $\alpha$ is a smooth function on $\I$. We further assume there is a functional relation 
between $\alpha$ and $\beta$,  
\ben \label{asscalar2}
\beta = W'(\alpha),
\een 
where $W$ is a smooth function of $\alpha$ with $W(0)=W'(0)=0$. 
\end{enumerate}

The prototype spacetime satisfying these asymptotic conditions (for any $W$) is pure AdS 
space with $\phi=0$ everywhere. In global coordinates its metric is given by
\ben\label{ads}
ds^2_0 = - (1 + r^2)\, dt^2 + \frac{dr^2}{1 + r^2}  
+ r^2 (d\theta^2   + \sin^2 \theta \, d\varphi^2),
\een
which can be brought into the 
form (\ref{metric1})-(\ref{metric2}) e.g. by choosing $\Omega = r^{-1}$. Thus, for pure AdS, a 
conformal completion can be obtained by taking 
$\tilde M$ to be the manifold obtained from $M$ by attaching the boundary $\I$ consisting of the 
points $\Omega=0$, and by taking the unphysical metric to be $d\tilde s_0^2 
= \Omega^2 ds^2_0$. One can likewise verify that the asymptotic conditions are also 
obeyed by the AdS-Schwarzschild and the AdS-Myers-Perry solutions, in which the scalar field vanishes.

If the above asymptotic conditions are combined with Einstein's equations, then one can obtain much 
more information on the asymptotic form of the metric near infinity. For instance, by using the 
expansion techniques of \cite{Hollands05} one finds 
\bena
\label{ds}
d\tilde s^2 &=& d\Omega^2 - \left( 1 + \frac{2 - \alpha^2}{4} 
\Omega^2 - \frac{4\alpha \beta}{9} \Omega^3
\right) dt^2 \nonumber\\
&+& \left( 1 - \frac{2 + \alpha^2}{4} \Omega^2 - \frac{4\alpha \beta}{9} \Omega^3
\right) d\sigma^2 \nonumber \\
&-& \frac{2}{3}\Omega^3 {\mathcal E}_{ab} \, dx^a dx^b + \dots, 
\eena
where $d\sigma^2 = d\theta^2 + \sin^2 \theta \, d\varphi^2$. 
The asymptotic expansion of the physical line element $ds^2$ is obtained by 
dividing the above expressions by $\Omega^2$.
The quantity ${\mathcal E}_{ab}$ is the leading order electric Weyl tensor, defined by 
\ben
{\mathcal E}_{ab}= \Omega^{-1} \tilde C_{acbd} \tilde n^c \tilde n^d. 
\een
This can be shown to be smooth at $\I$ if the field equations are satisfied, despite the inverse 
power of $\Omega$.

It is well known that with `reflective' boundary conditions, defined by $\a=0$, tachyonic scalars in AdS 
spacetime do not cause an instability provided their mass is above the BF bound. With $\a=0$ boundary 
conditions our theory (\ref{theory}) admits a positive energy theorem \cite{Abbott82,Gibbons83}, which ensures the 
total energy cannot be negative whenever this condition is satisfied.

In this paper we are concerned with a different class of boundary conditions, which we defined above by a function 
$W(\alpha)$. Although these boundary conditions are not in general invareiant under the full AdS symmetry group, they 
are invariant under global time translations so one can still define a conserved energy.
In \cite{Hertog04,Henneaux04,Hertog05} the conserved energy was constructed within the Hamiltonian framework of 
Henneaux and Teitelboim \cite{Henneaux85}. It was found that for all $W$ there is an additional 
surface term associated with the scalar field, which renders the conserved energy finite.
Here we study the stability of theories such as (\ref{theory}), with boundary conditions 
(\ref{metric2})-(\ref{asscalar2}). In particular, we wish to prove the conjecture of \cite{Hertog05}, 
which states that the energy is bounded from below whenever $W$ has a global minimum.

But first we present an alternative construction of the conserved charges in this context, in which we
adopt the covariant phase space formalism of Wald et al.~\cite{Wald00}. This provides a general 
algorithm for defining charges associated with symmetries preserving a given set of boundary conditions, and 
it will be useful when we come to analyze the positivity properties of the charges.

\subsection{Conserved Charges}

Asymptotic symmetries are diffeomorphisms $f$ of $\tilde M$ that preserve a prescribed set of 
boundary conditions. That is, if a solution $\Phi = (\phi, g_{ab})$ satisfies a given set of boundary 
conditions, then $f^* \Phi$ does, too. The asymptotic symmetries form an infinite-dimensional 
group. Of physical interest is the factor group $G = \mbox{Diff}(M)/\mbox{Diff}(M)_0$, 
where $\mbox{Diff}(M)_0$ is the subgroup of diffeos leaving a neighborhood of $\I$ pointwise invariant. 
For designer gravity boundary conditions defined by a generic $W$, 
the elements of $G$ can be identified with isometries of the Einstein
static universe. When $W(\alpha) \propto \alpha^3$, however, $G$ is larger and corresponds
to the {\em conformal} isometry group of the Einstein static universe, $G\cong O(3,2)$
\cite{Hertog04,Henneaux04}.
The Lie algebra of $G$ consists of vector fields $\xi^a$ tangent to $\I$ that Lie derive the 
metric of the Einstein static universe, modulo vector fields that vanish on $\I$. 
For simplicity, we will also refer to such vector fields as {\it asymptotic symmetries}.
We now carefully define the generators $\H_\xi$ on phase space associated with the asymptotic 
symmetries that leave invariant the above set of boundary conditions\footnote{Note that, since an 
infinitesimal asymptotic symmetry is only specified modulo vector fields on $\tilde M$ that vanish 
on $\I$, it follows that $\H_\xi$ can only depend on this equivalence class, i.e., it must 
vanish for any vector field that is zero on $\I$. This means, roughly speaking, that $\H_\xi$ 
cannot depend on derivatives of $\xi$ normal to $\I$.}. 

Consider first the variation of the Lagrange density ${\bf L}$. This can always be written in the form
\begin{equation}
\delta {\bf L} = {\bf F} \cdot \delta \Phi + d \bth, 
\end{equation}
Here, $\delta \Phi = (\delta g_{ab}, \delta \phi)$, the symbol ${\bf F}$ represents the field 
equations and $d\bth$ is the exterior differential of the $3$-form $\bth$ that
corresponds to the boundary term which would arise if the variation of ${\bf L}$ were performed 
under an integral sign. For the Lagrangian (\ref{theory}) it is given by 
\begin{equation}
\label{td}
\theta_{abc} = \frac{1}{2}(\nabla^d \delta g_e{}^e - \nabla^e \delta g_e{}^d 
+ 2\delta \phi \nabla^d \phi) \epsilon_{abcd}, 
\end{equation}
where $\mbox{\boldmath $\epsilon$} = d^4 x \sqrt{-g}$ is the volume form (identified with a $4$-form).
The antisymmetrized second variation $\bome$ of $\bth$ defines the (dualized) symplectic 
current, 
\begin{equation}
\label{dw}
\bome(\Phi; \delta_1 \Phi, \delta_2 \Phi) 
= \delta_1 \bth(\Phi; \delta_2 \Phi) - \delta_2 \bth(\Phi; \delta_1 \Phi) ,     
\end{equation}
so that $\bome$ depends on the unperturbed metric and scalar field and is skew in the pair of 
perturbations. 

The integral of the symplectic current over an achronal 3-dimensional submanifold $\Sigma$ 
defines the symplectic structure, $\sigma_\Sigma$, of the theory
\begin{equation}
\sigma_\Sigma(\Phi, \delta_1\Phi, \delta_2 \Phi) = 
\int_{\Sigma} \bome(\Phi; \delta_1 \Phi, \delta_2 \Phi) \, . 
\end{equation}
It follows from a general argument that if both perturbations satisfy the linearized equations 
of motion, then the symplectic current is conserved, $d\bome = 0$. From this one can deduce  
how $\sigma_\Sigma$ depends upon the choice of $\Sigma$. This goes as follows.
Let $\Sigma_1$ and $\Sigma_2$ two achronal surfaces ending on $\I$, which enclose 
a spacetime volume that is bounded by $\Sigma_1, \Sigma_2$, and the portion $\I_{12}$ 
of infinity bounded by the corresponding cuts $C_1$ and $C_2$ on which the surfaces end. 
By Stokes theorem, the difference 
between the symplectic form on both 3-surfaces is given by an integral
of $d\bome$ over the spacetime volume, plus the integral 
$\int_{\I_{12}} \bome$. However, for perturbations satisfying 
the linearized equations of motion, the symplectic current is given by
\bena\label{sympl}
\omega_{abc} \restriction \I &=& 
(\delta_1 \beta \delta_2 \alpha - \delta_2 \beta \delta_1 \alpha) \tilde \epsilon_{abc} 
\nonumber\\
&=& 
W''(\alpha) (\delta_1 \alpha \delta_2 \alpha - \delta_2 \alpha \delta_1 \alpha)  
\tilde \epsilon_{abc} = 0
\eena
on $\I$, where $\tilde \epsilon_{abc}$ is the integration element on $\I$ induced by 
$\tilde g_{ab}$, i.e., $4\tilde n_{[a} \tilde \epsilon_{bcd]} = \tilde \epsilon_{abcd}$.
Hence the symplectic structure $\sigma_\Sigma$ is independent of the choice of $\Sigma$. 
Note that this would not have been the case if $\alpha$ and $\beta$ had been two independent 
functions.

One would like to define the generator associated with a vector field 
$\xi^a$ representing an asymptotic symmetry by 
\ben\label{defcharge}
\delta \H_\xi = \sigma_\Sigma(\Phi; \delta \Phi, {\pounds_\xi} \Phi) \quad \forall \delta \Phi, 
\een
where $\Sigma$ is a partial Cauchy surface whose boundary, $S^2_\infty$, is a cut of $\I$.
This definition would imply that $\H_\xi$---if it exists--- generates (in the sense of Hamiltonian 
mechanics) the infinitesimal displacement (``Hamiltonian vector field'') 
$\delta \Phi = {\pounds_\xi} \Phi$, which in turn describes the action of an infinitesimal symmetry 
in the phase space of the theory. In addition, $\H_\xi$ would not depend on the cut $S^2_\infty$ 
chosen, since the symplectic form is independent of the choice of $\Sigma$. Hence, 
if one can indeed construct $\H_\xi$ in this way, it will be automatically 
{\it conserved}. Moreover, it also follows that the Hamiltonian vector fields associated with two 
infinitesimal symmetries $\xi^a, \eta^a$ satisfy the same algebra as ordinary vector fields on $M$ 
under the commutator~\cite{Hollands05}. 

To analyze the existence of $\H_\xi$ 
it is instructive, following~\cite{Wald00}, to first study the general structure of 
eq. (\ref{defcharge}). The right hand side of (\ref{defcharge}) can be written as \cite{Wald00}
\ben
\label{hdef}
\delta \H_\xi = \int_\Sigma \delta {\bf C}_a \xi^a 
+ \int_{S^2_\infty} [\delta {\bf Q}_\xi - \xi \cdot \bth], 
\een
Here, ${\bf C}_a$ are the constraints of the theory (identified with $3$-forms), 
and ${\bf Q}_\xi$ is the Noether charge, which is given in our case by
\ben
\label{nc}
\label{qdef}
Q_{ab}= -\frac{1}{2}(\nabla^c \xi^d) \epsilon_{abcd} \,. 
\een
Consistency requires $(\delta_1 \delta_2 - \delta_2 \delta_1) \H_\xi = 0$, so we must have
\ben
\label{consistency}
0 = 
\xi \cdot [\delta_2 \bth(\Phi, \delta_1 \Phi) - \delta_1 \bth(\Phi, \delta_2 \Phi)] = 
\xi \cdot \bome(\Phi; \delta_1 \Phi, \delta_2 \Phi)  
\een
on $\I$, or else $\H_\xi$ cannot exist. But this follows immediately from (\ref{sympl}). 
Here it enters there is a functional relation between $\beta$ and $\alpha$, illustrating
clearly that a complete set of boundary conditions includes a specification of $W$.

It follows from the consistency condition that there is a $2$-form ${\bf I}_\xi$ such that
\ben
\delta {\bf Q}_\xi - \xi \cdot \bth = \delta {\bf I}_\xi 
\een 
up to an exact form. We conclude that a solution to (\ref{hdef}) exists and is given by 
\ben
\label{hdef2}
\H_\xi = \int_\Sigma \xi^a {\bf C}_a + \int_{S^2_\infty} {\bf I}_\xi. 
\label{db}
\een
If the equations of motion are satisfied then the constraints ${\bf C}_a$ vanish identically
and $\H_\xi$ reduces to a surface integral. Finally, from (\ref{ds}), (\ref{td}) and (\ref{qdef}) 
it follows that a generator $\H_\xi$ that satisfies (\ref{hdef2}) is given explicitly by\footnote{
We thank Aaron Amsel for performing an independent check of~(\ref{charges}), 
and for filling in a gap in our original calculation.}
\ben \label{charges}
\H_\xi = -\int_{S^2_\infty} {\mathcal E}_{ab} t^a \xi^b \sqrt{\sigma} \, d^2 x 
- \int_{S^2_\infty} [W(\alpha) - \frac{1}{3} \alpha \beta] t^a \xi_a \sqrt{\sigma} \, d^2 x, 
 \een
where the cut at infinity has been chosen, without loss of generality, 
as the $S^2$ orthogonal to $t^a = (\partial/\partial t)^a$. Since $\H_\xi$ is 
defined through its variation, one can always add an arbitrary constant to (\ref{charges}). This 
constant is usually determined by requiring that $\H_\xi$ vanishes on exact AdS space
with $\phi=0$ everywhere, which holds for the above expression provided $W(0) = 0$. 

When $\xi^a = (\partial/\partial t)^a$, $\H_{\partial/\partial t}$ is the energy $E$ of the 
field configuration $(g_{ab}, \phi)$. 
The energy has a contribution from the electric Weyl tensor of the gravitational field, and 
an extra term that depends on the asymptotic profile of the scalar field and on the choice 
of boundary conditions defined by the function $W$. Note that the contribution from the scalar is 
absent for the generalized conformally invariant boundary conditions $\beta \propto \alpha^2$, as 
well as for the ``standard'' boundary conditions $\alpha = 0$. When $\xi^a =
(\partial/\partial \varphi)^a$ is a rotational Killing field on $S^2$, then 
$\H_{\partial/\partial \varphi}$ is the angular momentum $J$, and 
the expression for $\H_{\partial/\partial \varphi}$ coincides with the one obtained with standard boundary conditions.

The charges (\ref{charges}), which we defined by using the covariant phase space formalism of Wald et al. 
\cite{Wald00}, agree with the charges one obtains using the Henneaux-Teitelboim formalism 
\cite{Henneaux85}. Indeed, with generalized boundary conditions ({\ref{asscalar}) one finds the latter 
also acquire an explicit scalar contribution \cite{Hertog04,Henneaux04}. On the other hand, the definition of 
Ashtekar et al. \cite{Ashtekar84}, which is based purely on the electric Weyl tensor, obviously yields 
different charges in this context, which are in general not conserved. Finally, we expect that 
it should also be possible to construct the charges in designer gravity using the
``counterterm subtraction method''~\cite{Skenderis, Kraus}. Based on the general arguments
presented in~\cite{Marolf05}, we expect that the counterterm approach should yield charges 
that are equal to our charges (\ref{charges}), up to a constant offset. 

In the next section we compare our charges with the spinor charges~\cite{Gibbons83, Townsend84}, and use this 
relationship to analyze the positivity properties of the energy.

\setcounter{equation}{0}
\section{Lower Bound on the Energy}

The conserved energy $E=\H_{\partial/\partial t}$ cannot satisfy a positive energy theorem for all 
choices of boundary conditions, since for certain $W(\alpha)$ there exist globally smooth solutions 
with negative energy. Indeed, in \cite{Hertog05b} it was shown that our theory (\ref{theory}) with 
$\beta = -k\alpha^2+\epsilon \alpha^3$ boundary conditions (with $k \gg \epsilon >0$) admits e.g. negative energy spherical solitons. Furthermore, with AdS-invariant boundary conditions 
$\beta = -k \alpha^2$ one can even construct smooth solutions that have arbitrary negative energy
\cite{Hertog04b}. This raises the question what (if any) choices of boundary conditions $W$ in designer 
gravity yield a theory with a stable ground state. In this section we address this point. In
particular, we show there is a lower bound on the energy $E$ of any solution that satisfies boundary 
conditions for which $W$ has a global minimum. The central idea of our proof is to relate the 
Hamiltonian charges (\ref{charges}) to the spinor charges, which can be shown to be manifestly positive for all $W$.

First consider a future directed timelike vector field $\xi^a$ for which asymptotically
\ben
\xi^a = 
\left( \frac{\partial}{\partial t} \right)^a + \omega \left(
\frac{\partial}{\partial \varphi }\right)^a, 
\quad  {\rm on}\,\, \I, 
\een
where $-1<\omega<1$ is a real constant. Assume that $M$ admits a spin structure, and
let $\psi$ be a spinor field on $M$ such that asymptotically 
$\overline \psi \gamma^a \psi = \xi^a$. Let ${\bf B}$ be the 2-form
with  components
\ben\label{2form}
{B}_{ab} = 
\frac{1}{2}(\overline \psi \gamma^{[c} \gamma^d \gamma^{e]} \widehat \nabla_e \psi 
+ {\rm h.c.})\epsilon_{abcd} \, , 
\een
where $\gamma_a$ are the curved space gamma matrices and
\ben
\widehat \nabla_a \psi 
= \left[ \nabla_a - \frac{1}{2}\gamma_a \cosh(\phi/\sqrt{2}) \right]\psi. 
\een
The spinor charge is defined by 
\ben\label{spinorcharge}
{\mathcal Q}_\xi = \int_{S^2_\infty} {\bf B}.
\een 
The following identity~\cite{Townsend84} allows one to prove its positivity:
\ben\label{spinid}
(d{\bf B})_{ijk} = [2(\widehat D_l \psi, \widehat D^l \psi) + |\lambda|^2] \epsilon_{ijk} - 2|
\gamma^l \widehat D_l \psi|^2 \epsilon_{ijk} 
\een
with
\ben
\lambda = \frac{1}{\sqrt{2}} \gamma^a (\nabla_a \phi) \psi +  \sinh (\phi/\sqrt{2}) \psi \, . 
\een
Here $i,j, \dots$ denote indices tangent to a spacelike 3-surface $\Sigma$ spanning the $S^2_\infty$ at 
infinity, $\epsilon_{ijk}$ is the induced volume 3-form, and $\widehat D_i$ is the derivative 
operator obtained from $\widehat \nabla_a$ by projecting the tangent space index into $\Sigma$. 
Finally, the hermitian inner product $(\lambda,\lambda)$ is defined by 
$\overline \lambda \gamma^a \lambda u_a$, where $u^a$ is the unit timelike normal to $\Sigma$.
The first term in $d{\mathbf B}$ is a manifestly non-negative density. 
The second term can be negative, but it can be set to zero by requiring $\psi$ to satisfy
the {\it Witten condition},
\ben
\gamma^i \widehat D_i \, \psi = 0. 
\een
In Appendix A we prove that our choice of boundary conditions indeed allows one to impose this 
condition, and at the same time require that asymptotically $\overline \psi \gamma^a \psi = \xi^a$. 
Assuming that $\Sigma$ has no boundary components other than $S^2_\infty$ (such as inner
boundaries ending on the horizon of a black hole), 
it now follows from Stokes theorem that $\Q_\xi = \int_\Sigma d{\mathbf B} \ge 0$. 

Now define $\tilde \psi = \Omega^{1/2} \psi$ and $\tilde \gamma_a = \Omega \gamma_a$, which are 
smooth functions on the boundary $\I$, and write the 2-form $B_{ab}$ as 
$\Omega^{-2} \tilde B_{ab}$. The 2-form $\tilde B_{ab}$ is given by the same expression 
(\ref{2form}), but with all quantities replaced by their corresponding tilded quantities and 
with $\widehat \nabla_a$ replaced by $\Omega^{1/2} \widehat \nabla_a \Omega^{-1/2}$. From the 
asymptotic expansion (\ref{expansion}) of the spinor field $\psi$, and the asymptotic
expansions of the metric and the scalar field $\phi$, eqs. (\ref{ds}) and (\ref{asscalar}), 
it follows that
\bena
\label{1010}
\Omega^{1/2} \widehat 
\nabla_a (\Omega^{-1/2} \tilde \psi) 
&=& \Omega^2 \left[ \frac{1}{2} {\mathcal E}_{ab} 
\tilde \gamma^b 
+ \frac{1}{12} \alpha \beta \tilde \gamma_a \right] \tilde \psi \nonumber\\
&+& \dots \, , 
\eena
Here the dots stand for terms which do not contribute to the 
spinor charges (see also Appendix A). 
Inserting this formula into the expression 
(\ref{spinorcharge}) for the spinor charge yields the following relation between the Hamiltonian 
charges $\H_\xi$ and the spinor charges ${\mathcal Q}_\xi$,
\ben
\label{hq}
\H_\xi = {\mathcal Q}_\xi - \int_{S^2_\infty} W(\alpha) t^a \xi_a \sqrt{\sigma} \, d^2 x. 
\een
Since the spinor charge is positive and since $\H_\xi = E + \omega J$, where $J$ is the 
angular momentum, this gives
\ben
E + \omega J \ge \int_{S^2_\infty} W(\alpha) \sqrt{\sigma} \, d^2 x \, ,
\een
and consequently
\ben\label{bound}
E \ge 4\pi \, {\rm inf} \, W + |J|.
\een  
Thus the energy is bounded from below in designer gravity for all asymptotic conditions (\ref{asscalar2}) that are 
defined by a function $W(\alpha)$ which is bounded from below.

Note also that, unlike $\H_\xi$, the spinor charge $\Q_\xi$ is not conserved in general. 
Instead it depends on the choice of the cross section $S^2_\infty$ at infinity. This follows 
immediately from the fact that it differs from $\H_\xi$ by the term 
$\int_{S^2_\infty} W(\alpha)$, which 
depends on the choice of the cross section, because $\alpha$ is, in general, time dependent. 

Finally, we note that for designer gravity theories that have a dual field 
theory description, the bound (\ref{bound}) is obvious from the point of view 
of the dual field theory.
Indeed, it is known that all configurations in the dual CFT with $W=0$ 
satisfy $E >= |J|$. The change in the energy
by the deformation is $\int_{S^2_\infty}  <W(O(x))> \sqrt{\sigma} d^2 x$. 
This leads in the
large $N$ limit - which corresponds to the supergravity approximation we 
used - to $\int_{S^2_\infty} W(<O>) \sqrt{\sigma} d^2 x$, 
which clearly leads to
(\ref{bound}). The above result, which only involved arguments in the
gravity theory, can therefore be viewed as a test of the AdS/CFT 
correspondence.

\setcounter{equation}{0}
\section{Ground State}

Since we have established that the energy $E$ is bounded from below if $W$ has a global minimum,
the question arises what are the solutions $\Phi_0 = (g_{ab}, \phi)$ that minimize $E$. 
It is clear from (\ref{bound}) that if $W$ is everywhere positive, the ground state of our theory 
with boundary conditions $\beta = W'$ remains pure AdS spacetime with $E=0$. On the other hand if $W$ has negative 
regions the lower bound (\ref{bound}) leaves open the possibility of negative energy solutions which, 
if they exist, would render pure AdS nonlinearly unstable.

For certain $W$, the solitons discussed in \cite{Hertog05} provide explicit examples of negative energy 
solutions. These solitons are regular, static, spherically symmetric solutions with line element
\ben
ds^2=-h(r)e^{-2\chi(r)}\, dt^2+h^{-1}(r)\, dr^2+r^2 \, d\sigma^2. 
\een
The set of soliton solutions of our theory (\ref{theory}) can be labeled by the value of $\phi$ at the origin. 
For each $\phi(0)$ and with regular boundary conditions at the origin, one can (numerically) integrate 
the field equations outward and get a soliton. Asymptotically, $\phi$ behaves like (\ref{asscalar}), 
so one obtains a point in the $(\alpha,\beta)$ plane. Repeating for all $\phi(0)$ yields a curve 
$\beta_s(\alpha)$ where the subscript indicates this is associated with solitons. We plot this 
curve\footnote{Since $V(\phi)$ is even, it suffices to consider positive $\phi(0)$ 
corresponding to positive $\alpha$.} in Fig.1.

For a given $W(\alpha)$, the allowed solitons are simply given by the points 
where the soliton curve intersects the boundary condition curve: $\beta_s(\alpha) = W'(\alpha)$.
For example, with boundary conditions $\b(\a)=-\a^2+.22 \a^3$, which were discussed in 
\cite{Hertog05b}, one finds precisely one soliton with negative energy, $E_s \approx -20\pi$.
\begin{center}
\begin{figure}[htb]
\begin{picture}(0,0)
\put(0,100){$\beta$} 
\put(350,220){$\alpha$}
\end{picture}
\mbox{\epsfxsize=12cm \epsfysize=8cm \epsffile{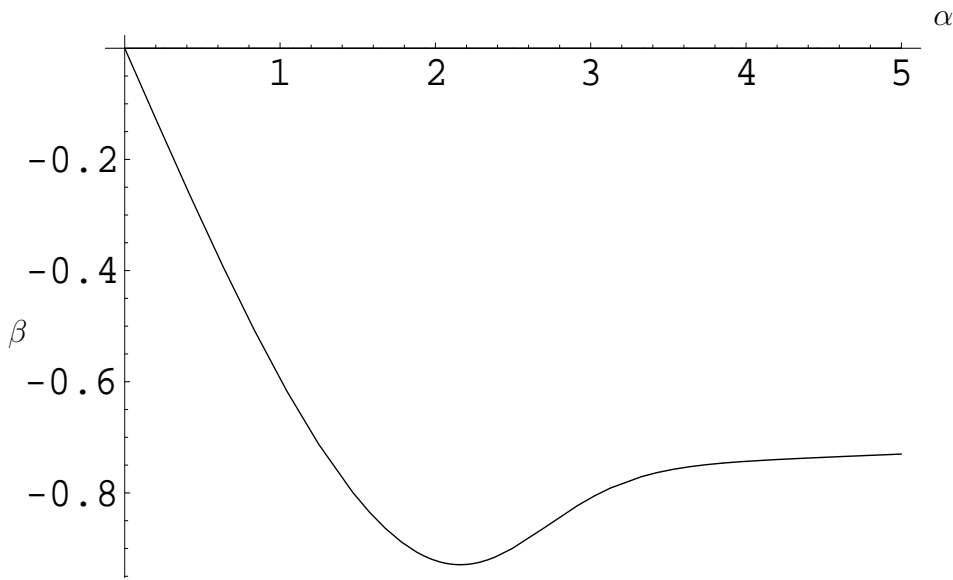}}
\caption{The function $\beta_s$  obtained from the solitons.} 
\label{1}
\end{figure}
\end{center}

The AdS/CFT correspondence \cite{Maldacena98} relates the solitons to nontrivial vacua in 
the dual field theory, which lives on the boundary $\I$. With $W=0$ boundary conditions, the field 
theory that is dual to $D=4$, ${\cal N}=8$ gauged supergravity is the 2+1 conformal field theory (CFT) 
describing the low energy excitations of a stack of M2-branes. The scalar $\phi$ that we have retained 
corresponds to a dimension one operator $\O$ \cite{Aharony98}. Imposing more general boundary conditions 
(\ref{asscalar}) corresponds to 
the addition of a potential term $\int W({\cal O})$ to the CFT action \cite{Witten02}. According to the standard 
AdS/CFT dictionary, the solitons determine the expectation value $\langle {\cal O} \rangle$ in different 
{\it vacua} of the deformed dual field theory. In particular, given a soliton with $\b_s =W'$, one has 
$ \langle {\cal O} \rangle = \a_s$. Furthermore, from the soliton curve $\b_s(\a)$ one can construct the following 
function,
\ben\label{effpot}
{\cal V}(\alpha) = -\int_{0}^{\a} \b_{s} (\tilde \a) d\tilde \a + W(\alpha),
\een
which captures this correspondence between solitons and field theory vacua in a nice way.
Indeed it can be shown that for any $W$ the location of the extrema of ${\cal V}$ yield the vacuum expectation values
$ \langle {\cal O} \rangle = \a$, and that the value of ${\cal V}$ at each extremum yields
the energy of the corresponding soliton \cite{Hertog05}.

The dual AdS/CFT interpretation of the solitons naturally leads one to conjecture \cite{Hertog05} that 
if ${\cal V}$ has a global minimum, the true ground state of the theory with $\b=W'$ boundary conditions 
should be either the lowest energy soliton (if $E_s <0$) or pure AdS. In the previous section we have shown that the 
conserved energy is indeed bounded from below if $W$ (and therefore ${\cal V}$) has a global minimum. However - 
although the dual field theory argument is suggestive - this does not immediately imply that the lowest energy state is
one of the solitons. This is especially so because the lowest energy soliton does not saturate the lower bound 
(\ref{bound}). Instead the actual soliton energy contains an additional strictly positive contribution, which equals 
the spinor charge. 
More generally, it follows from Fig.1 there are $W$ with a negative global minimum (and $W(0)=W'(0)=0$) which admit 
no soliton solutions at all. The AdS/CFT interpretation of this is that the corresponding (deformed) field theories 
have no vacua with $\langle {\cal O} \rangle \neq 0$, suggesting pure AdS remains the true ground state of the bulk
theory. In fact, our numerical calculations show that unless the global minimum of $W$ is sufficiently negative, all 
solitons have positive energy (and hence correspond to metastable field theory vacua). These considerations suggest it 
should be possible to strengthen the bound (\ref{bound}).

We have been able to make partial progress towards establishing the nature of the ground state
in theories where (\ref{bound}) provides a nontrivial (negative) lower bound on the energy.
If one assumes that the asymptotic value $\alpha_0$ of the scalar field in the ground state $\Phi_0$
is time-independent\footnote{Numerical calculations provide some support for this claim. We have compared, 
for several $W$, the energy of solitons to the energy of $O(4)$-invariant instantons, which continue to spherical 
Lorentzian solutions with $\frac{\partial}{\partial t} \alpha_0\neq 0$. In all cases we find the instantons 
have larger energy than the lowest energy soliton.}, $\frac{\partial}{\partial t} \alpha_0=0$, one can apply a 
method of Wald
and Sudarsky \cite{Sudarsky92} to our theory, to show that $\Phi_0$ necessarily has to be {\em static}, 
i.e., that there exists a globally defined time-coordinate $t$ such that the vector field 
$t^a = (\partial/\partial t)^a$ Lie-derives the solution, ${\pounds}_t \Phi_0 = 0$ (and therefore is in 
particular a surface-orthogonal Killing field). We refer the reader to Appendix B for the details of this 
argument. It essentially consists of proving that if $\Phi_0$ is a solution to Einstein's 
equations, then one can construct a linearized perturbation $\delta \Phi$ satisfying the linearized 
equations of motion and the linearized boundary conditions, such that $\delta E(\Phi_0) < 0$, unless 
$\Phi_0$ is static. A local minimum of the energy by definition has $\delta E(\Phi_0) = 0$. Hence it 
must be static. 
It remains an open problem, however, to show that the static ground state $\Phi_0$ must be 
spherical, and therefore be the lowest energy soliton. In particular, the fact that we only know the 
soliton solutions numerically makes it rather difficult to apply the techniques of Masood-ul-Alam~\cite{Masood93}.

\setcounter{equation}{0}
\section{Generalizations}

Although we have concentrated on a single scalar field with $m^2=-2$ in four dimensional  
$\N=8$ supergravity, 
our arguments apply to more general theories. Consider, for example, gravity in $d$ spacetime 
dimensions coupled to a scalar field that takes values in an $N$-dimensional Riemannian target space 
manifold  $(X,G_{IJ})$, with standard kinetic term and potential $V$. Assume that $V$ arises from a 
superpotential $P$,
\ben
V = -8(d-1) P^2 + 8(d-2) G^{IJ} P_{, I} P_{, J}\, . 
\een
and that $V$ has a critical point $\phi_0 \in X$ where $V(\phi_0) <0$. This means one can define standard AdS 
boundary conditions for which the theory admits a positive energy theorem.

However, if some of the scalar masses $m^2_I$  [the eigenvalues of the 
matrix $V_{,I}{}^J(\phi_0)$] are negative and within the range 
(\ref{range}), then
one can consider alternative asymptotic conditions that are defined by a smooth function $W:X \to {\mathbb R}$.
As before, one requires there be a conformal factor $\Omega$ such that $\tilde g_{ab} = \Omega^2 g_{ab}$ is 
smooth at $\I$ and given by eq.~(\ref{metric2}), and one further requires that\footnote{The asymptotic 
conditions are modified \cite{Hertog04,Henneaux04} for scalar fields which saturate the BF bound, such that 
$\lambda_+ = \lambda_- \equiv \lambda$.} 
\ben
\phi^I = \phi_0^I + \Omega^{\lambda_{-}^{I}} \alpha^I + \Omega^{\lambda_{+}^{I}} 
\beta^I + \dots, 
\een
where $\beta_I = W_{, I}$, 
\ben\label{roots}
\lambda^{I}_{\pm} = {d-1 \pm \sqrt{(d-1)^2 + 4 m^2_{I}}\over 2} \, , 
\een
and where we have assumed without loss of generality that coordinates in field space 
have been chosen so that 
$G_{IJ}(\phi_0) = \delta_{IJ}$ and $V_{,IJ}(\phi_0) = m^2_I \delta_{IJ}^{}$.
As in the case of a single scalar field, one can again demonstrate that the consistency condition 
(\ref{consistency}) holds, and hence that the generators for the asymptotic symmetries can be
consistently defined. The asymptotic symmetry group is generically the 
isometry group of the $(d-1)$-dimensional Einstein static universe, but for 
$W = \sum k_I (\alpha^I)^{(d-1)/\lambda_-^I}$ it is the conformal group $O(d-1,2)$.
The charges ${\mathcal H}_\xi$ must be given by conformally invariant expressions 
in that case. By analogy with the 4-dimensional case discussed above, 
we therefore expect that they are given by  
\ben
\label{conservedd}
\H_\xi = -\int_{S^{d-2}_\infty} {\mathcal E}_{ab} t^a \xi^b \sqrt{\sigma} \, d^{d-2} x 
- \int_{S^{d-2}_\infty} \left[ W
- \frac{1}{d-1} \sum_I \lambda_{-}^I \alpha^I \beta^I \right] t^a \xi_a \sqrt{\sigma} \, d^{d-2} x \, , 
\een
where ${\mathcal E}_{ab}$ is the leading order electric part of the Weyl tensor given now by
\ben
{\mathcal E}_{ab} = \frac{1}{d-3} \Omega^{3-d} \tilde C_{acbd} \tilde n^c \tilde n^d \, .
\een 
In particular, it would follow that the 
energy is finite. It may be verified by comparing ${\mathcal H}_\xi$ to the 
spinor charge that it would satisfy 
a similar lower bound as before, eq.~(\ref{bound}).
A more detailed discussion will appear elsewhere
\cite{Amsel05}.

Finally let us mention that 
the condition $\beta_I = W_{,I}$ is not the most general requirement ensuring
that the consistency condition~(\ref{consistency}) is satisfied. In fact, 
the consistency condition is equivalent to the requirement that the 
points $(\alpha^I, \beta_J) \in {\mathbb R}^N \times {\mathbb R}^N$
lie on an $N$-dimensional submanifold $S \subset {\mathbb R}^N \times {\mathbb R}^N$ with 
the property that the pull-back of the 2-form $d\beta_I \wedge d\alpha^I$ to $S$ vanishes
(such submanifolds are called ``Lagrangian''). Locally, one can always find coordinates
$(\alpha^I, \beta_J)$ and a function $W$ such that $S$ is defined by 
$\beta_I = W_{, I}$, but this need not be the case globally. 
It would be interesting to carry out our analysis for general Lagrangian 
submanifolds.

\setcounter{equation}{0}
\section{Discussion}

We have studied the stability of designer gravity theories in asymptotically AdS spacetimes, where one couples 
gravity to a tachyonic scalar (with $m^2$ in the range (\ref{range})) which obeys boundary 
conditions that are defined by an essentially arbitrary function $W(\a)$.
We first derived Hamiltonian generators of the asymptotic symmetries in theories of this type. A comparison of the
resulting charges with other definitions of charges in asymptotically AdS spaces showed that the generators differ 
in general from the spinor charges, except when $W=0$. The latter charges are thus not conserved, but their 
positivity, even in designer gravity, enabled us to establish a lower bound (\ref{bound}) on the 
conserved energy of any solution that satisfies boundary conditions for which $W(\a)$ has a global minimum. 
This proves that a large class of designer gravity theories have a stable ground state.

In cases where $W$ has a negative global minimum, the lowest energy spherical soliton is a natural candidate for the 
true ground 
state of the theory. The dual field theory description of certain supergravity theories in particular provides strong
evidence for this. It is more difficult, however, to establish the nature of the ground state using purely 
gravitational arguments. We have only obtained a partial result on this, showing that solutions locally
minimizing the energy---and thus in particular the global minimum of the energy---have 
to be static. One problem is that at present the solitons 
are only known numerically. A more profound issue is that with boundary 
conditions that admit negative energy solutions one should not expect the 
lower bound (\ref{bound}) to be saturated by the ground state of the theory. This is 
because (\ref{bound}) does not take in account the contribution from the spinor charge, 
which is positive and vanishes only in perfect AdS spacetime. 
It is therefore possible that our bound can be further strengthened, and
perhaps even generalized. 

Indeed, motivated by the AdS/CFT correspondence, \cite{Hertog05} conjectured there would be a lower bound on the energy 
in all designer gravity theories where the function ${\cal V}(\a)$, given in eq.(\ref{effpot}), has a global minimum. 
It is clear from Fig.1, however, that for ${\cal V}(\a)$ to be bounded from 
below it is sufficient that $\b_{s} <W' $ at large $\a$, which includes certain $W$ that are unbounded\footnote{The 
dual field theory description suggests that even $W \sim -\a^3$ boundary conditions yield a well-defined theory 
\cite{Hertog05b}. However, the corresponding ground state in the bulk, if it exists, is likely to be a classically 
singular spacetime.}. To establish a lower bound on the energy for this class of boundary conditions no doubt requires
a better handle on the value of the spinor charge in the true ground state of the theory.

\bigskip

\centerline{{\bf Acknowledgments}}
\bigskip

This work was supported in part by NSF grants PHY-0244764 and PHY-0354978
and DOE grant DE-FG02-91ER40618. We would like to thank 
W.~Junker and E.~Schrohe, as well as A.~Amsel and D.~Marolf 
for discussions. 

\appendix
\setcounter{equation}{0}
\section{Witten Spinor Condition}

We now prove that there exist globally defined, smooth 
 asymptotically Killing spinors satisfying the Witten 
condition \cite{Witten81}
\ben\label{witten}
L\psi \equiv \gamma^i \widehat{D}_i \, \psi = 0, 
\een
on a smoothly embedded spacelike 3-surface $\Sigma \subset M$. We also prove 
that $\psi$ has a suitable asymptotic 
expansion near infinity $S^2_\infty = \partial \Sigma$, and explicitly
display the first non-trivial coefficient of that expansion, which is 
used in the calculations reported in the main text. 
We assume that $\Sigma$ is so that it can be covered by 
a finite collection of conical regions, and that its boundary meets $\I$ transversally, 
implying in particular that the induced metric $h_{ij}$ on $\Sigma$ is asymptotically
hyperbolic. We also assume that Einstein's equation hold for $(M,g_{ab},\phi)$.

Our strategy for constructing the desired solution to the Witten equation is to first obtain an 
approximate solution to the Witten equation with the desired asymptotics. This is a smooth spinor 
field satisfying eq.~(\ref{witten}) with a source, $L \lambda = J$, where 
$J$ vanishes to all order in $\Omega$ near $S^2_\infty$. An exact solution $\psi$ 
will then be obtained by constructing a suitable solution $\mu$ to the equation
$L\mu = -J$, and setting $\psi = \lambda + \mu$. 

We seek the approximate solution $\lambda$ as an asymptotic series in $\Omega$, 
\ben
\label{asympt}
\lambda \sim \psi_0 + \Omega \psi_1 + \Omega^2 \psi_2 + \dots + \Omega^N \psi_N + \dots \, ,
\een
where $\psi_0$ is a Killing spinor in exact AdS-space (i.e., $\widehat \nabla_a \psi_0 = 0$ in 
exact AdS with vanishing $\phi$) such that asymptotically
\ben
\xi^a = \overline \psi_0 \gamma^a \psi_0^{}
\een
for a given (timelike or null) 
asymptotic symmetry $\xi^a$. The spinor fields $\psi_1, \psi_2, \dots$ are determined
recursively by the condition that $\lambda$ be an asymptotic solution to the Witten equation. The
recursion relations are obtained as follows: One first rewrites the Witten equation in terms of the 
derivative operator associated with the unphysical metric $\tilde g_{ab} = \Omega^2 g_{ab}$, the 
unphysical gamma matrices $\tilde \gamma_a = \Omega \gamma_a$, and the quantities 
$\tilde \psi_j = \Omega^{1/2} \psi_j$, which are required to be smooth at $S^2_\infty$. 
Then the unphysical metric and 
derivative operator are expanded around the corresponding unphysical metric of exact AdS space as in
eq.~(\ref{ds}). One finally expands the resulting equation in powers of $\Omega$ and equates terms 
with equal powers. If this is done, then one finds that all spinor fields $\psi_1, \psi_2,
\dots$ are uniquely determined in terms of $\psi_0$. Furthermore, if $\{ \alpha_k \}$ is 
an appropriately fast decreasing sequence of positive real numbers and $\chi$ is 
a smooth compactly supported function on $\mr$ which is equal to $1$ in a neighborhood of 
$\Omega = 0$, then the series
\ben
\lambda = \sum_{k=0}^\infty \chi(\Omega/\alpha_k) \Omega^k \psi_k
\een
converges, and defines a globally smooth function on $\Sigma$, 
and $\tilde \lambda = \Omega^{1/2} \lambda$ is smooth at $S^2_\infty$. It satisfies 
the Witten equation with a source $J$ vanishing to all orders in $\Omega$. 
Clearly, the smooth function $\lambda$ has the asymptotic expansion~(\ref{asympt}), 
the first terms of which are given explicitly by
\bena \label{expansion}
\tilde \psi_1 &=&  0 \nonumber \\
\tilde \psi_2 &=&  \frac{1}{16} \alpha^2 \tilde \psi_0 \nonumber \\
\tilde \psi_3 &=&  -\frac{1}{6} {\mathcal E}_{bc} \tilde h_a{}^b 
\tilde \gamma^a \tilde 
\gamma^c \tilde \psi_0 + \frac{1}{36} \alpha \beta \tilde \psi_0 \nonumber \\
&& + \frac{\sqrt{2} -3}{192} \alpha^2 \tilde P_- \tilde U + 
\frac{1}{96} ( \tilde \gamma^d \tilde \nabla_d \alpha^2 - t_b 
\tilde \gamma^b t^d \tilde \nabla_d \alpha^2 ) \tilde \psi_0 \, ,
\eena 
where $\tilde U$ is a spinor field determined by $\tilde \psi_0$ 
that is smooth at scri, and 
where $\tilde P_\pm = (1\pm \tilde n^a \tilde \gamma_a)/\sqrt{2}$.
This expansion is used in obtaining eq.~(\ref{1010}).  

We next obtain from the asymptotic solution $\lambda$ an exact solution $\psi=\lambda+\mu$ of the 
Witten equation by constructing a suitable solution $\mu$ to $L\mu = -J$. The existence 
of such a solution will be established with the help of the following key lemma.
 
\noindent
\paragraph{\bf Lemma:} Let $(M, g_{ab}, \phi)$ be a solution of the equations of motion satisfying 
our asymptotic conditions. Let $u$ be any smooth spinor field on $\Sigma$ of compact support, and 
let 
\ben
\| u \|^2 = \int_\Sigma [(\widehat D_i u,\widehat D^i u) +
(1+r^2)^{-1} |u|^2 ]\sqrt{h} \, d^3 x, 
\een
where $r$ is a radial coordinate equal to $\Omega^{-1}$ near $\partial \Sigma =
S^2_\infty$.
Then there exists a positive constant independent of $u$ such that
\ben
\c^{-1} \| u \|^2 
\le \int_\Sigma |Lu|^2 \sqrt{h} \, d^3 x 
\le \c \| u \|^2 
\een
\paragraph{\it Proof:} We begin with the first inequality. We cover $\Sigma$ with a finite number of
open cones $\Gamma$, each of which is generated by the orbits of a ``radial'' vector field $r^i$. 
We normalize this vector field so that $r^i D_i r = 1$. For each such cone, we can define a 
hermitian, positive definite bilinear form $b(u, v)$ in the spinor bundle, which depends smoothly on 
the base point, and which is covariantly constant along the integral curves of $r^i$ in the sense 
that
\ben
r^i D_i [b(u, v)] = b(r^i \widehat D_i u, v) + b(u, r^i \widehat D_i v). 
\een 
Such a hermitian form can be defined e.g. by choosing an arbitrary positive definite hermitian form 
in the spinor fiber over the apex of $\Gamma$, and then by parallel transporting this form along the
integral curves of $r^i$ with the connection $r^i \widehat D_i$. From the asymptotic expansion of 
the metric it is then possible to establish that there is a constant such that 
\ben\label{ineq}
\c^{-1} (1+r^2)^{-1/2} b(u,u) \le |u|^2 \le \c(1+r^2)^{1/2} b(u,u)
\een
at every point in $\Gamma$. Consider now the identity 
\ben
0 = \int_0^\infty r^i D_i [r b(u,u)] \, dr = 2\int_0^\infty r \, {\rm Re}[ 
b(u, r^i \widehat D_i u)] \, dr + \int_0^\infty b(u,u) \, dr \, .  
\een
Integrating this relation over $\widehat \Gamma$ (the quotient of
$\Gamma$ by its generators) with the surface element $d^2 \sigma$
of the 2-sphere, using the Cauchy-Schwarz inequality, and using 
the inequality $r^2 (r^i w_i)^2 \le \c \, h^{ij} w_i w_j$ for some 
constant and all $w_i$, we obtain
\ben
\int_\Gamma b(u,u) \, dr d^2 \sigma \le \c \int_\Gamma b(\widehat D^i
u, \widehat D_i u) \, dr d^2 \sigma. 
\een
We now use the geometrical inequality (\ref{ineq}), together with the fact 
that the measure $(1 + r^2)^{1/2} dr d^2 \sigma$ is equivalent to the
measure $\sqrt{h} \, d^3 x$ on $\Gamma$. This yields the following inequality,
\ben
\label{hardy}
\int_\Gamma (1+r^2)^{-1} |u|^2 \sqrt{h} \, d^3 x \le \c \int_\Gamma
(\widehat D^i u, \widehat D_i u)\sqrt{h} \, d^3 x
\een
for some constant. Since $\Sigma$ is the union of finitely many cones 
$\Gamma$, the above inequality holds also also for the the entire 
3-manifold $\Sigma$. 

So far we have only used the asymptotic behavior of metric and the scalar field, but not the fact 
that they satisfy Einstein's equations. We will now use Einstein's equations to derive the first 
inequality in the lemma from the inequality~(\ref{hardy}). When Einstein's equations are satisfied, 
then eq.~(\ref{spinid}) holds. Integrating this equation over $\Sigma$ for some compactly 
supported spinor field $u$ yields
\bena
\int_\Sigma |L u|^2 \sqrt{h} \, d^3 x
&=&  \int_\Sigma (\widehat D^i u, \widehat D_i u) \sqrt{h} \, d^3x \\
&+& \frac{1}{2} \int_\Sigma
\left| \frac{1}{\sqrt{2}} \gamma^a (\nabla_a \phi) u + \sinh (\phi/\sqrt{2})u
\right|^2 \sqrt{h} \, d^3 x \,\, . \nonumber 
\eena
Estimating the second term on the right side by 0 from below, and combining the resulting 
inequality with~(\ref{hardy}) then yields the first half of the inequality in the lemma. 

To obtain the second half of the inequality claimed in the lemma, we 
estimate the second term on the right side instead by 
\ben
\left| \frac{1}{\sqrt{2}} \gamma^a (\nabla_a \phi) u + \sinh (\phi/\sqrt{2})u
\right|^2 \le \c (1+r^2)^{-1} |u|^2
\een
from above, which in turn is an immediate consequence of our
asymptotic conditions on the metric and the scalar field.//

We now use the lemma to establish a solution to the equation $L \mu = -J$. We 
first use the lemma to establish the solution of a distributional solution $\mu$. Let $F$ be the 
linear functional on $C^\infty_0$ defined by 
\ben
F(u) = -\int_\Sigma (J, u)
\sqrt{h} \, d^3 x \, . 
\een
Then, by the Cauchy-Schwartz inequality, and the first half of the inequality in the lemma, we get 
\bena
|F(u)| &\le& \left( \int_\Sigma |u|^2 (1+r^2)^{-1} \sqrt{h} \,
d^3 x\right)^{1/2} \left( \int_\Sigma |J|^2 (1+r^2) \sqrt{h} \,
d^3 x\right)^{1/2} \nonumber \\
&\le& {\rm const.} \left( \int_\Sigma |L u|^2
  \sqrt{h} \, d^3 x \right)^{1/2} \, .
\eena
We interpret this inequality as saying that $F$ is a bounded functional in the scalar product given 
by the right side of this inequality. Let $H$ be the Hilbert space defined by this inner product 
(which, by the second inequality of the Lemma is identical with the Hilbert space obtained from 
$\|\, \cdot \, \|$). By the Riesz representation theorem, there is hence an element $v \in H$ 
such that $F(u) = \int (Lu, Lv) \sqrt{h} \, d^3x$
for all $u \in C^\infty_0$. Again by the inequality in the lemma, every element in $H$ is in 
particular locally square integrable, so in particular a distribution on $\Sigma$. Thus, $v$ is 
a solution in the distributional sense of the equation $L^\dagger L v = -J$. 
Hence, $\mu = L^\dagger v$ is the desired distributional solution to $L\mu = -J$, and 
$\psi = \lambda + \mu$ is a global solution to the Witten equation.

It remains to prove that $\psi$ is smooth, and that it has an asymptotic expansion, i.e., can 
be approximated by ($\Omega^{-1/2}$ times) 
a polynomial in $\Omega$ with smooth coefficients, up to 
a function on $\Sigma$ which is smooth everywhere including the 
boundary, and vanishes to any desired order in $\Omega$. Note that, 
although this is true by construction for the spinor field $\lambda$, this has not 
been shown to be the case yet for $\mu$, since it follows from our construction 
only that $\mu \in L^2(\Sigma, \sqrt{h}\, d^3 x)$. However, the machinery developed in 
\cite{mazzeo} will now allow us to establish that $\mu$ is smooth on $\Sigma$ including 
infinity, and that it has an asymptotic expansion. 
Following this formalism, we associate with $L$ a ``normal operator'' 
$N_q(L)$ for each $q \in S^2_\infty$. This is a differential operator 
acting on $(y^1, y^2, \Omega) \in T_q S^2_\infty \times \mr_+$, 
which captures the asymptotics of the 
operator $L$ near the point $q$. It is given by 
$N_q(L) = \Omega^{3/2} N_+ \Omega^{-1/2}$, where
\ben
N_\pm = \Gamma \partial_\Omega + D_{\mr^2} \mp \frac{3}{2}\Omega^{-1}(1 \mp \Gamma) \, ,
\een
and where $D_{\mr^2} = \gamma^1 \partial_{y^1} + \gamma^2 \partial_{y^2}$ is the ordinary
Dirac operator on $\mr^2$, and where $\Gamma = i\gamma^1 \gamma^2$. 
The dependence upon $q$ has dropped out in the present case because $L$
is spherically symmetric asymptotically. One next determines for which $\delta \in \mr$ the 
operator $N_q(L)$ has a nontrivial kernel in the 
space $\Omega^\delta L^2(\Omega^{-3} \, d\Omega
d^2 y)$ (the integration element 
reflects the large distance behavior of the volume element $\sqrt{h} \, d^3 x
\sim \Omega^{-3} \, d\Omega d^2 \sigma$).
This kernel may be determined e.g. by noting that any element in its kernel 
necessarily has to be $\Omega^{1/2}$ times an element 
in the kernel of $N \equiv N_- N_+$, which in 
turn may be identified with the matrix operator
\ben
N = \left(
\begin{array}{cc}
\partial_\Omega^2 + 3\Omega^{-1} \partial_\Omega - |k|^2 & 0\\
0 & \partial_\Omega^2 + 3\Omega^{-1} \partial_\Omega - |k|^2 - 3\Omega^{-2}
\end{array}
\right) \, ,
\een
acting on functions of $\Omega \in \mr_+$, where we have also 
performed a Fourier transform in $y^1, y^2$. The  
equation $N(\chi_1, \chi_2) = 0$ can be solved in terms of Bessel functions. 
It turns out that 
$\chi_1$ must be a linear combination $a_1\Omega^{-1} I_{1}(|k|\Omega) +
a_2\Omega^{-1} K_{1}(|k| \Omega)$, 
while $\chi_2$ must be a linear combination $b_1 \Omega^{-1} I_2(|k|\Omega) +  
b_2\Omega^{-1} K_2(|k|\Omega)$. 
Thus, no element in the kernel of $N_q(L)$ 
can be in $\Omega^{\delta} L^2(\mr_+, \Omega^{-3} \, d\Omega)$ for $\delta > -2$, 
in particular,  $N_q(L)$ is injective on $L^2(\mr_+, \Omega^{-3} d\Omega)$. Since
$L\mu = -J$, $\mu \in L^2(\Omega^{-3}\, d\Omega d^2 \sigma)$, and since $J$ 
has an asymptotic expansion near the boundary, it now follows by 
Prop.~7.17 of~\cite{mazzeo}, that also $\mu$ is smooth and that it 
has an asymptotic expansion, with leading 
coefficient $\Omega^{-1/2}$. Consequently, 
also $\psi = \lambda + \mu$ has such an expansion, which is what we desired to show. 
Actually, since the expansion of $\psi$ is unique (and hence equal to that of $\lambda$ by
construction), the expansion of $\mu$ must in fact vanish, i.e., $\mu$ must in fact vanish 
to all orders in $\Omega$. 

\setcounter{equation}{0}
\section{Staticity of Extrema of Energy}

In this Appendix we show that if the asymptotic value $\alpha_0$ of the scalar field is 
time-independent, i.e. $\frac{\partial}{\partial t} \alpha_{0}=0$, then solutions $\Phi_0 = (g_{ab}, \phi)$ that minimize 
the energy $E=\H_{\partial/\partial t}$ must be {\em static}, i.e., there exists a globally defined
time-coordinate $t$ such that the vector field $t^a = (\partial/\partial t)^a$ Lie-derives the 
solution, ${\pounds}_{\partial/\partial t} \Phi_0 = 0$ (and therefore is in particular a surface-orthogonal Killing 
field). 
We prove this by applying a method of Wald and Sudarsky \cite{Sudarsky92} to our theory. 
This method proceeds by showing that, if $\Phi_0$ is a solution to Einstein's equation, then one can
construct a linearized perturbation $\delta \Phi$ satisfying the linearized equations of motion and
the linearized boundary conditions, such that $\delta E(\Phi_0) < 0$, unless $\Phi_0$ is static. A 
local minimum of the energy by definition has $\delta E(\Phi_0) = 0$. Hence it must be static. 

In order to construct the desired perturbation, we make the additional technical assumption that 
our energy minimizing asymptotically AdS solution $\Phi_0 = (\phi, g_{ab})$ is such that the metric
admits a foliation by spacelike maximal 3-surfaces $\Sigma$ 
(i.e., surfaces with vanishing trace $K$ of the extrinsic curvature), 
meeting infinity $\mathcal I$ transversally. 
The desired perturbation can be constructed as follows. 
Pick a maximal surface $\Sigma$, let $h_{ij}$ and $K_{ij}$ be the induced 
metric and extrinsic curvature, and let $\phi$ and $p$ be the scalar field and its normal derivative
on $\Sigma$. Assuming that our theory has a well-posed initial value formulation, these can be 
identified with initial data for the solution $\Phi_0$. We now consider initial data for the 
linearized solution $\delta \Phi$ of the form
\begin{equation}\label{linsol}
\delta K_{ij} = -6fK_{ij} - K_{ij}, \,\,
\delta h_{ij} = 4fh_{ij}, \,\,
\delta p = -6fp-p, \,\,
\delta \phi = 0, 
\end{equation}
where $f$ is a function on $\Sigma$ to be determined. Equations (\ref{linsol}) correspond to 
a linearized solution on $M$ if and only if $f$ is chosen such that the linearized constraint
equations $\delta {\mathbf C}_a=0$ are satisfied. Under our assumption $K=0$, these are
equivalent to the equation
\ben
\label{elli}
D_i D^i f - \mu f = \rho, 
\een
where 
\ben
\mu = K_{ij} K^{ij} + p^2 + 2 + 
\cosh(\sqrt{2} \phi), \quad \rho = \frac{1}{4} K_{ij} K^{ij} + \frac{1}{4} p^2. 
\een
Equation~(\ref{elli}) is a Laplace equation for the asymptotically hyperbolic metric
$h_{ij}$, with a potential and a source. 
It follows from $\frac{\partial}{\partial t} \alpha_0=0$ that the potential satisfies 
$\mu = 3+O(r^{-2})$ and that the source satisfies $\rho = O(r^{-6})$. 
One can establish using the methods of \cite{mazzeo,mazzeo1} that such an equation 
(\ref{elli}) has an everywhere smooth solution $f$ which asymptotically behaves like 
$r^{-3} f_0$, with $f_0$ a smooth function on $\partial \Sigma = 
S^2_\infty$, the cut at infinity. In particular, the 
corresponding linearized perturbation $\delta \Phi$ will then satisfy the linearized asymptotic 
conditions. In order to calculate the change in the energy associated with this linearized 
perturbation, we note that due to the rapid fall-off of $f$ at infinity, $\delta \Phi$ has a
vanishing $\delta \alpha$. Hence, the change in energy for the linearized perturbation comes 
entirely from the Weyl-tensor term in eq. (\ref{charges}), and is given by 
\bena
\delta E(\Phi_0) 
&=& -\delta \int_{S^2_\infty} {\mathcal E}_{ab} t^a t^b
\sqrt{\sigma} \, d^2 x \nonumber \\
&=& -\lim_{r \to \infty} \delta \int_{S^2_r} r^5 \Bigg( {\mathcal R}_{ij} - K_{ik}
  K^k{}_j - \frac{1}{3} p^2 h_{ij} \nonumber \\
&& \quad - \frac{1}{6} h_{ij} D^k \phi D_k
  \phi - \frac{1}{3} h_{ij}V - \frac{1}{2} D_i \phi D_j \phi
\Bigg) r^i r^j \sqrt{\sigma} \, d^2 x \nonumber \\
&=& 24 \int_{S^2_\infty} f_0 \sqrt{\sigma} \, d^2x \,\, , 
\eena
where in the second line, we have used the Gauss-Codacci equations and Einstein's equation, and 
where ${\mathcal R}_{ij}$ is the Ricci tensor of $h_{ij}$. Because $\rho, \mu \ge 0$, the elliptic
equation (\ref{elli}) is of a form to which the Hopf maximum principle~\cite{Hopf} applies, that is, within each
domain $D \subset \Sigma$ the solution $f$ must assume its maximum on $\partial D$, or it must be 
constant. However, $f$ vanishes at infinity, so $f \le 0$ everywhere. Consequently, also 
$f_0 \le 0$, meaning that $\delta E(\Phi_0) \le 0$. Moreover, we will now argue that, if 
$\rho \neq 0$, then also $f_0 \neq 0$, and so we must have in fact that $\delta E(\Phi_0) < 0$, 
meaning that the perturbation causes the energy to decrease (to first order). Consequently, in order
to avoid a contradiction with $\delta E(\Phi_0)=0$ for minimizing solutions, it must be true that 
$\rho = 0$ and therefore that 
\ben
p=0, \quad K_{ij} = 0
\een
everywhere on the maximal 3-surface $\Sigma$. Since $M$ 
can be foliated by such surfaces, this implies that the solution $\Phi_0$ is globally static. 
Thus, in order to finish the proof, we must show that $f_0 = 0$ implies $\rho = 0$. 
The proof of this is accomplished by the following lemma.
\medskip

\noindent
\paragraph{\bf Lemma:}
Let $f$ be a smooth solution to $D^i D_i f - \mu f = \rho$, where $\mu$ and $\rho$ are smooth, {\em nonnegative}
functions on $\Sigma$ satisfying $\mu = 3 + O(r^{-2})$ and $\rho = O(r^{-2})$, and where $D_i$ is the derivative 
operator of an asymptotically hyperbolic metric $h_{ij}$.
Assume that $f = O(r^{-3-\epsilon})$
for some $\epsilon > 0$. Then $\rho = 0=f$.   
\medskip

\noindent
{\em Proof:}
We can write the asymptotically hyperbolic metric on $\Sigma$ as 
\ben
h_{ij} \, dx^i dx^j = \Omega^{-2}(d\Omega^2 + k^2 d\sigma^2), 
\een
where $\Omega =  O(r^{-1})$ is a non-negative defining function of the surface $S^2_\infty$ at infinity, and 
$k$ is a smooth function of $\Omega$ and the angles such that $k \to 1$ and $k' \to 0$ as 
$\Omega \to 0$. Prime denotes a derivative with respect to $\Omega$. Consider now the quantity
\ben
G(\Omega) = -\Omega^{-1} \int_{S^2_\Omega} k^{-2} f \sqrt{\sigma} \,
d^2 x \, .
\een
By the maximum principle, we have $G \ge 0$, and from the assumption about the asymptotics of $f$, we know that 
$\Omega^{-2-\epsilon} G(\Omega)$ is continuous as $\Omega \to 0$, for some $\epsilon > 0$. We will 
now show that there must be an $\Omega>0$ for which $G$ vanishes. This will finish the proof, for 
then we must have that $f(x)=0$ at some interior point $x \in \Sigma$. Indeed, the Hopf maximum 
principle leaves only the alternatives that the global maximum of $f$ (equal to 0) must be attained 
on the boundary $\partial \Sigma$, or that $f$ vanishes identically. Thus, the second
alternative must be the case, implying that also $\rho=0$ identically. 

To prove that $G=0$ for some $\Omega > 0$, we integrate the equation for $f$ over $S^2_\Omega$, and 
use that $f \le 0, \rho \ge 0$ one obtains an evolution equation for $G$ of the form
\ben
(\Omega^{-1}(\Omega G)')' \le 3\Omega^{-2} G + (\Omega H)' + F \, \, , 
\een
where 
\bena
H(\Omega) &=& -2\Omega^{-2} \int_{S^2_\Omega} k^{-2} (\log k)' f \sqrt{\sigma}
\, d^2 x \\
F(\Omega) &=& -\Omega^{-1} \int_{S^2_\Omega} k^{-2} \xi f \sqrt{\sigma} \, d^2 x \,\, .
\eena
where $\xi = \Omega^{-2}[p^2 + K_{ij}
K^{ij} + 2 \sinh^2(\phi/\sqrt{2})] \ge 0$ is smooth as $\Omega \to
0$. We now integrate the inequality from $0$ to $\Omega$, using the
estimate
\ben
\int_0^\Omega x^{-2} G(x) \, dx \le \left( \sup_{0 \le x \le \Omega}
x^{-2-\epsilon} G(x) \right) 
\int_0^\Omega x^\epsilon \, dx 
= \frac{1}{1+\epsilon} \Omega^{1+\epsilon} \gamma(\Omega)
\een
for the first term, where we defined
\ben
\gamma(\Omega)=\sup_{0 \le x \le \Omega}
x^{-2-\epsilon} G(x) \, .
\een
We furthermore use the estimate
\ben
\Omega H(\Omega) \le \left( \sup_{0 \le x \le \Omega}
x^{-1} |(\log k)'(x)| \right) \Omega^{3+\epsilon}
\gamma(\Omega) \equiv \Omega^{3+\epsilon} c_1(\Omega) \gamma(\Omega)
\een
for the second term, and the estimate 
\bena
\int_0^\Omega F(x) \, dx &\le& \left( \sup_{0 \le x \le \Omega} \xi(x) \right) 
\left( \sup_{0 \le x \le \Omega}
x^{-2-\epsilon} G(x) \right) \int_0^\Omega x^{1+\epsilon} \, dx
\nonumber\\
&\equiv& \Omega^{2+\epsilon} c_2(\Omega) \gamma(\Omega)
\eena
for the third term on the right side. Note that all three functions
$c_1, c_2, \gamma$ are non-negative, continuous at $\Omega= 0$, 
and monotonically increasing. From these estimates, we 
get the inequality
\ben
(\Omega G)' \le \frac{3}{1+\epsilon} \Omega^{2+\epsilon}
\gamma(\Omega) + c_1(\Omega) \Omega^{3+\epsilon} \gamma(\Omega) + 
c_2(\Omega) \Omega^{4+\epsilon} \gamma(\Omega) \, .
\een
Now pick a fixed $\Omega_0 > 0$, and integrate one more time from $0$
to $\Omega \le \Omega_0$. Then one obtains the inequality
\ben
\Omega^{-2-\epsilon} G(\Omega) \le \left(
\frac{3}{(1+\epsilon)(3+\epsilon)} 
+ \frac{1}{4+\epsilon} c_1(\Omega_0) \Omega_0 
+ \frac{1}{5+\epsilon} c_2(\Omega_0) \Omega_0^2 \right)
\gamma(\Omega_0) \, .
\een
maximizing the left side over all $\Omega \le \Omega_0$ then gives the
final inequality
\ben
\gamma(\Omega_0) \le \left(
\frac{3}{(1+\epsilon)(3+\epsilon)} 
+ \frac{1}{4+\epsilon} c_1(\Omega_0) \Omega_0 
+ \frac{1}{5+\epsilon} c_2(\Omega_0) \Omega_0^2 \right)
\gamma(\Omega_0) \, .
\een
It is now clear that if $\Omega_0$ is chosen small enough, then the
first term on the right side will dominate the other terms, 
and we get an inequality of the 
form $\gamma(\Omega_0) \le (1-\delta) \gamma(\Omega_0)$, where
$\delta >0$. This is possible only if in fact $\gamma(\Omega_0)=0$ and
hence $G(\Omega_0)=0$, which is what we wanted to show.//


\end{document}